# When Children Program Intelligent Environments: Lessons Learned from a Serious AR Game

Evropi, E.S., Stefanidi

University of Bremen, Bremen, Germany, & Institute of Computer Science, Foundation for Research and Technology – Hellas (FORTH), Heraklion, Crete, Greece, evropi@uni-bremen.de

Maria, M.K., Korozi

Institute of Computer Science, Foundation for Research and Technology – Hellas (FORTH), Heraklion, Crete, Greece, korozi@ics.forth.gr

Asterios, A.L., Leonidis

Institute of Computer Science, Foundation for Research and Technology – Hellas (FORTH), Heraklion, Crete, Greece, leonidis@ics.forth.gr

Dimitrios, D.A., Arampatzis

Institute of Computer Science, Foundation for Research and Technology – Hellas (FORTH), Heraklion, Crete, Greece, arabatzis@ics.forth.gr

Margherita, M.A., Antona

Institute of Computer Science, Foundation for Research and Technology – Hellas (FORTH), Heraklion, Crete, Greece, antona@ics.forth.gr

George, G.P., Papagiannakis

Institute of Computer Science, Foundation for Research and Technology – Hellas (FORTH), Heraklion, Crete, Greece, & Computer Science Department, University of Crete, Heraklion, Greece, papagian@ics.forth.gr

While the body of research focusing on Intelligent Environments (IEs) programming by adults is steadily growing, informed insights about children as programmers of such environments are limited. Previous work already established that young children can learn programming basics. Yet, there is still a need to investigate whether this capability can be transferred in the context of IEs, since encouraging children to participate in the management of their intelligent surroundings can enhance responsibility, independence, and the spirit of cooperation. We performed a user study (N=15) with children aged 7-12, using a block-based, gamified AR spatial coding prototype allowing to manipulate smart artifacts in an Intelligent Living room. Our results validated that children understand and can indeed program IEs. Based on our findings, we contribute preliminary implications regarding the use of specific technologies and paradigms (e.g. AR, trigger-action programming) to inspire future systems that enable children to create enriching experiences in IEs.

## 1 INTRODUCTION

In recent years, the emergence of the concept of Intelligent Environments (IEs), (i.e. spaces including artifacts made intelligent with computing power [1], that proactively, but sensibly, support people in their daily lives [2]) and pertinent paradigms (i.e. Internet of Things (IoT) [3]), has driven advancements in various technological domains. This has led to an abundance of systems and technologies, which initially began as research prototypes, but are nowadays offered -in the form of smart devices and artifacts- as commercial products that can be easily integrated into common living environments (e.g. homes, offices, schools, museums). In addition to their functionality, these systems incorporate sophisticated mechanisms that permit end-users to configure and control them according to their personalized needs and preferences. This feature has resulted in a proliferation of research work towards determining appropriate ways for end-users to easily program such devices to support their everyday lives. An example is trigger-action programming, used to prescribe the users' own logic towards manipulating the smart artifacts in their surroundings [4].

These efforts mainly targeted adults (belonging in the age group of 25-50), who usually buy, install and use IoT devices in their homes, as the relevant demographics demonstrate [5–8]. According to literature [9–11], most adults are likely to have a child within that age range; therefore, their children, especially the younger ones (aged 5-12), are expected to be living in IEs and use smart artifacts before entering adolescence [12]. However, there is a pertinent gap in the relevant literature, since children are not usually considered to be end-users of IEs, let alone of systems allowing their programming. Similar to adults, children have diverse interests and preferences, while they also expect a certain degree of control over what is going on in their own homes [13, 14]; therefore, it is important to enable them to customize their surroundings by manipulating (in a child-friendly manner) various smart devices that regard them. Despite the fact that it is already established that young children are capable of programming [15, 16], and regardless of the identified importance of engaging young learners in technology and engineering education [17–19], limited research has addressed the issue of equipping children with tools that enable them to program IEs. As a result, it has yet to be determined whether younger children have the ability to comprehend what an IE is and what it offers, and if they can actually program the smart artifacts it includes.

Beyond the acquisition of engineering and computational-thinking skills, the programming of IEs has the potential to empower children in their everyday activities and cultivate various skills towards their development. Particularly, encouraging children to actively participate in the management of their intelligent surroundings can enhance feelings of responsibility, independence, and the spirit of cooperation, similarly to when they help around with other household activities [20, 21]. Furthermore, previous research has demonstrated that children develop numerous important skills during that age; diSessa [22] and Papert [23] have argued that programming constitutes an empowering, creative, and fertile medium for children's activities, while studies with Logo [24] have proved that it can help children improve visual memory and cultivate number-sense, as well as develop problem-solving techniques and language skills [25]. Moreover, according to Piaget's theory of cognitive development [26], children of 7-11 ages develop their logical thought [27], while their intelligence "is demonstrated through logical and systematic manipulation of symbols related to concrete objects" [26]. Additionally, several studies from the '80s until today [28–31], have concluded that children construct much of their knowledge through active manipulation of the environment; in particular, children aged 5-8 rely on active manipulation of real artifacts to connect abstract material, such as ideas and statements, to something observable and imaginable [32]. Based on the above, it can be argued that since programming IEs encompasses manipulation of physical objects, task organization and rational thinking, then it could offer a wide spectrum of benefits for child development.

To this end, this paper reports the findings of a study with fifteen children aged 7-12, who were asked to play four levels of progressive difficulty of a serious Augmented Reality (AR) game, allowing the creation of trigger-action rules that dictate the behavior of their surroundings, using tablets. The main goals of the study were to understand whether children comprehend the potential of IEs and what they can offer, and if they can indeed create the rules that define the behavior of the environment's artifacts. The results demonstrate that children understand the concept and can indeed program IEs, which is in line with previous positive research [16] towards children developing computational skills through interaction with appropriate programming games. In particular, our findings support that in intelligent domestic environments, one potential form of interaction with technology can be AR games that transform the environment into a "coding playground" [33], where children (aged 7-12), can safely explore its capabilities. At the same time, they can sharpen their computational skills, through programming their surroundings to either meet their needs or just for fun. Moreover, we found that combining these concepts with trigger-action programming delivers a child-friendly system for end-user programming of IEs, since children from the age of 5 are already familiar with verbal conditional statements [34], which makes the formation of trigger-action rules an intuitive process. Based on these findings, we discuss preliminary implications that can inspire future systems aiming to serve as tools that permit children to create enriching experiences in IEs. Finally, it should be noted that such approaches are particularly timely, as the current social distancing measures resulting from the COVID-19 pandemic have increased the interest in alternative, supplementary, and/or complimentary ways to learn new skills, such as programming, in different modalities (e.g. games and AR).

## 2 BACKGROUND THEORY & RELATED WORK

### 2.1 Children & Programming

Computer programming has been acknowledged as an important competence for the development of higher-order thinking, in addition to algorithmic problem-solving skills [22, 24]. Once thought of as an arcane craft practiced by techies, coding is now recognized by educators and theorists as a crucial skill and a new literacy for children [35]. Computational thinking has become the umbrella term encapsulating what computer science has to contribute to reasoning and communicating in today's ever-increasing digital world [35]. The work in [12, 36] highlights the importance of introducing programming and coding at a very young age. In particular, at around 5-6 years old, kids can learn the basic concepts of coding, such as algorithms, sequences, decomposition, and loops [37]. Between the age of 5 and 8, children advance from the preoperational to the concrete operational stage [16], where they rely on physical symbols and representation, and as they grow up, they are better able to take others' perspectives, use logic-based causal reasoning, and rely less directly on physical representations of ideas [38–41]. Thus, the importance of being able to create connections between physical and virtual objects in the age range of 7-12 becomes apparent.

Various paradigms have been utilized to allow children to program. Back in 2009, Eisenberg [42] demonstrated that new programming materials and physical settings will shift the nature of children's programming towards a more informal, approachable, and natural activity. Visual programming environments enable children to create complex programs with little training [43, 44], which has been proven to increase motivation in educational activities [45, 46]. Moreover, block-based environments improve learnability for novices [47] and provide the ease-of-use of direct manipulation interfaces [48]. Scratch is the most prominent web-based example. It is an educational language [49] employing a building-block metaphor for creating programs like a puzzle, which has demonstrated in practice that visual programming is appealing to kids [50]. Aiming to harness the appeal of VR, VR-OCKs [51] is designed for children and teenagers to learn basic programming concepts, by commanding the game character's movements via a block language. Regarding mobile devices, Catroid [52] is a noteworthy block-based environment allowing novice users -starting from the age of 8- to develop their own animations and games.

Tangible programming is also beneficial for children, since it can ease the learning of complex syntax [53] and has a positive impact on learning, situational interest, enjoyment, and programming self-beliefs [54]. Notable examples are AlgoBlocks [55], Tern [53], and Bots and (Main)Frames [54], which all use a building-block metaphor. Another example is T-maze [56], where children create maze maps and complete maze-escaping tasks using wooden programming blocks and sensors. In Strawbies [57], children use wooden tiles to construct programs in front of an iPad, to guide the main character on its quest. In the domestic context, Media Cubes [58] allows consumer devices to be controlled by tangible programming elements, while in robotics, KIBO [59] allows children aged 4-7 to control the movement of a robot through wooden building blocks.

At the same time, spatial ability and thinking play critical roles for children to build expertise in STEM (Science, Technology, Engineering, and Mathematics) [60, 61]. Thus, various educational applications have been developed, harnessing the benefits of AR towards delivering interactive experiences seamlessly interwoven with the surrounding environment [62, 63]. For instance, HyberCubes [64] is an AR platform for children that offers an introduction to basic computational thinking skills, by spatially orienting handmade paper cubes. Moreover, AR Scratch [65] adds AR functionalities to the Scratch platform, allowing users to see virtual objects in AR through a camera, and to control the virtual world through physical interaction with special markers. Augmented Creativity [66] employs both visual programming and AR to enable robot programming, while RoboTIC [67] is a serious AR game for learning programming using Microsoft HoloLens, which employs visual metaphors of roads and traffic signs. Finally, AR-maze is a tangible programming tool using AR, allowing children to create their own programs by arranging programming blocks and debug or execute the code with a mobile device [68].

## 2.2 Programming Intelligent Environments

Physical interactive environments can come in many forms, such as NYU's Immersive Environments [69], MIT's KidsRoom [70], and the University of Maryland's StoryRooms [71]. In particular, when these spaces include IoT-enabled devices, they are called Intelligent Environments. The enabling technologies that support these computationally enhanced physical environments can be found in Ubiquitous Computing [72], IoT [73], Ambient Intelligence [74], Augmented Reality [75], tangible bits [76], and graspable UIs [77]. The proliferation of IEs -as a result of the increased availability of various kinds of smart devices- creates the need for simple yet effective ways of managing them. Providing end-users with tools suitable for programming their environment is not new [78–80]. Trigger-action programming is considered the most straightforward approach of end-user programming [81] (i.e. people who are not professional software developers being able to instruct software artifacts to exhibit automated behavior), available in many shapes (Conversational Agents [82], graphical authoring tools [83, 84], etc.). For instance, Seraj et. al [85] presented a block-based educational programming tool in a web-based environment, allowing users to learn the general purpose of programming and create applications in the context of Intelligent Environments.

However, to the best of our knowledge, children's perception, understanding, and reaction towards IEs have not been addressed in literature, while only very few approaches exist permitting children to program IEs. Story-Room Kits [86] describes programming environments that permit young children to build interactive story spaces using sensors and actuators to augment everyday objects. Additionally, the system presented in [87] offers an educational application to program smart homes for older children (12-13 years old). Utilizing Google's Blockly, it provides its users with a block-based visual programming interface, and effectively allows them to program by themselves. However, these approaches do not extensively investigate whether younger children (7-12 years) actually comprehend the concept of IEs, or whether they can successfully program them.

## 3 GAME DESCRIPTION

A gamified AR spatial coding prototype called *MagiPlay* was used as a tool to permit the participants of this study to program an IE. It provides children with a fun way to program their surroundings, while at the same time enhancing their computational thinking skills. MagiPlay has already been evaluated in terms of usability via a user-based evaluation experiment (N=10), which assessed its overall functionality [88]. Using their tablets, children can move around their surrounding environment and collect the smart artifacts they wish to program (i.e. IoT-enabled objects) by pointing at them with the device's camera in AR; upon successful detection, the identified items can be used as the "ingredients" of a rule. The Rule Editor takes the form of a virtual baseplate, superimposed -via AR- over a flat surface like a coffee table, where the players can build rules by connecting virtual 3D LEGO-like bricks on top of it (Figure 1).

AR technology was the selected vehicle towards the realization of MagiPlay's objectives for multiple reasons. Firstly, it is recognized to bring many benefits regarding children's education [89] such as: (i) increased content understanding, (ii) long-term memory retention, (iii) increased engagement and motivation, and (iv) improved collaboration, which in turn insinuate that such a technology can effectively provide amusing and attractive ways for children to develop computational literacy. Additionally, AR increases task-related intuitiveness [90] and ensures location awareness [91]; the latter is of great significance in our context, since children are asked to discover and collect (via AR) smart artifacts from their physical surroundings. Following the same line of thought, the overall artifact collection process assists children in better understanding which artifact's behavior they are going to manipulate, by allowing them to make direct connections between the real, physical objects and the 3D virtual blocks representing them; thus, AR bridges the virtual world of programming with the real world of triggers-actions and artifacts. In any case, younger children (5-8 years old) benefit from having active experiences with manipulatives that promote the development of associations between concrete and symbolic representations [92].

In addition to the use of AR technologies, the inherent concept of programming a smart space by itself motivates children, as they can witness first-hand the results of their programming activities [87]. Moreover, fun and engagement are ensured through MagiPlay's game-like nature and the integration of proper techniques, like those proposed in gamification theory [93]. MagiPlay includes levels that children need to complete in order to advance in the game, while they also acquire experience points that reflect their overall in-game progress (e.g. number and complexity of created rules). As for the use of the "building-block" (bricks) metaphor, it should be mentioned that playing with bricks brings many advantages, such as learning to recognize shapes and colors and developing manual skills [94].

Regarding the rule creation process, research has shown that rule-based programming systems make it easier for children to create working programs [95]. Accordingly, MagiPlay follows the trigger-action paradigm [96], enabling children to create simple trigger-action rules to connect services, devices, and artifacts, and determine their behavior. Every rule is composed of one or more triggers, and one or more actions, following the structure: **when** TRIGGER(S) **then** ACTION(S). The trigger refers to an event that might occur, such as "the door opens", while an action constitutes the step that should be taken in response to that event, e.g. "the light should turn on". Despite the limitations stemming from using simple trigger-action rules, which seem to hinder the expressivity of the programs that can be created, according to [97] "a task specific language with appropriate tool support provides an ideal environment for users to create their own applications".

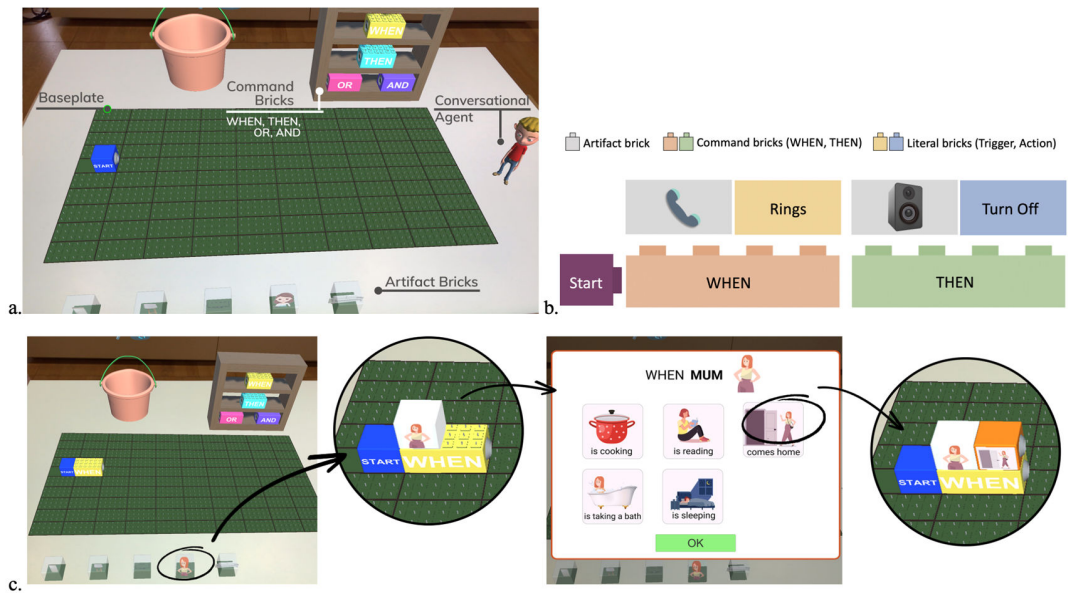

Figure 1: a) Snapshot of the game, b) Creating the rule "WHEN the Phone rings, THEN the Speakers turn off", c) Snapshot of the game while creating the trigger "WHEN Mum comes home", © Icons Designed by {pikisuperstar, brgfx, pch.vector} / Freepik

MagiPlay offers players different types of bricks: a) **command bricks**, representing the coding commands that can be used within the trigger-action programming principle (i.e. WHEN, THEN, AND, OR), available on a virtual wooden Library (Figure 1a); b) **artifact bricks**, representing the smart objects, the user roles, and the individual residents of the IE, visible in the front of the baseplate; and c) **literal bricks**, which are generated when an artifact brick is connected to a command brick, and represent either the exact value that will trigger the condition of the rule (e.g. door "opens") or the action that the artifact will take (e.g. "turn on" the light). When building a rule, command bricks can be placed next to each other, while the connection between a command and an artifact brick is established by placing the artifact on top of the command brick (Figures 1b, 1c). For example, the rule "*when the phone rings, the speakers should turn off*" is represented by a pair composed of an artifact brick (phone) and a literal brick ("Turn Off") on top of a command brick (WHEN), which in turn is connected with another command brick (THEN) that bears an artifact (speakers) and a literal ("Turn Off") brick.

Regarding gameplay, players progress through the game levels by carrying out simple programming tasks. At every subsequent level, the complexity of the expected rule(s) increases, while additional artifacts may become available. In order to create a rule, children move freely around and scan using their devices' camera their surroundings. When a smart artifact is detected, if they choose to collect it, its respective artifact brick will be displayed on the table. A level is completed once the player has collected all the necessary artifacts and has appropriately combined them with command bricks to form a number of rule(s) on the baseplate. Context-sensitive help is provided via a Conversational Agent in the form of a 3D animated character, aiming to facilitate the rule creation process by helping players either on-demand (e.g. the child asks which services are provided by an artifact) or automatically, by inferring when a player is "confused" (e.g. a mistake is repeated, or the child remains idle for a long time). In case of errors, the agent outlines the problem to the player (e.g. "You are missing an artifact block..."), so that they can address it. When a created rule is syntactically correct, the player is awarded the respective experience points based on its complexity, the included artifacts get unlocked (i.e. they can be used in other rules as well) and the rule can be simulated and deployed.

Simulation of a rule can be achieved by interacting: either with the physical objects of the environment (e.g. opening the door), or with the artificial objects of the virtual environment (e.g. open the door virtually through the tablet device), or with a combination of the two. In more detail, the following combinations are available: (i) virtual-to-virtual: artificially trigger a rule and view its effects virtually through the screen, (ii) virtual-to-physical: virtually trigger a rule and view its effects on the physical environment or (iii) physical-to-virtual: physically manipulate devices to trigger a rule and view its effects virtually through the screen. As an example of the first (i) case, the player can tap either on the collected digital TV artifact, or the real TV objects in AR, and "virtually" turn it on from the respective menu option that appears next to it, and notice the lights in their virtual environment dim, without this affecting the physical environment. After verifying that the rule has the desired effect, the player can immediately deploy it. Regarding deployment, a rule can be set to run once, or for a specific period (e.g. for the next 3 hours), or always. As soon as a rule is deployed, the IE integrates it, and from that point on is able to adapt its behavior accordingly when the rule gets triggered.

## 4 METHODOLOGY

We conducted a study (N=15) with the participation of children aged 7-12, in order to observe children interacting with the block-based, gamified AR spatial coding prototype and the IE, and note their comments and reactions, so as to understand whether they comprehend the concept and can successfully program the IE. In particular, this age group was selected considering Piaget's theory of cognitive development, where children aged 7-12 belong in the concrete operational stage and thus are capable of concrete problem-solving [98]. The study and its protocol were approved by the Ethics Committee of our research center (Reference Number: 57/18-11-2019) and were designed following guidelines for testing with children described in [99] and practices presented in [100]. Given that it involved the participation of minors, their legal guardians/parents gave their informed consent for inclusion before participation. The study took place in the simulation space of an "Intelligent Living Room" [101], which is a part of an "Intelligent Home" facility. The "Intelligent Living Room" is a room furnished and decorated like a typical living room, giving the impression of a welcoming and familiar space, rather than a "sterile" research lab [102, 103]. The "Intelligent Living Room" artifacts that were used in the study were two smart tabletop lamps, a smart TV, a smart door, a smart sofa, an IP phone, and a smart speaker. The goal of the study was to answer the following questions:

Q1. *Can children (aged 7-12) comprehend the concept of IEs?* Our objective was to investigate whether children are able to comprehend the concept of IEs, in the sense that an IE is able to monitor the surroundings and the individual users in it, anticipate user needs and act accordingly to facilitate them with their daily activities. In particular, we aimed to assess their understanding regarding the opportunities offered by IoT-enabled objects and the fact that users can create the rules that define their behavior on-demand.

Q2. *Can children (aged 7-12) dictate the behavior of an IE by creating appropriate rules?* We wanted to examine if children are able to find the appropriate smart objects, whether they can understand their programmable features, and if they can create the conditions that lead to a specific result (form a sequence that comprises a rule).

A total of fifteen children took part in the study. Our primary selection criteria were to include a wide range of ages (Figure 2) within the targeted age group (M=9.5, SD=1.78), and gender balance (7 girls, 8 boys). The initial participants' list was lengthier since thirty more children were invited through personal contacts of the research team (classmates of their children); however, due to the COVID-19 outbreak, some of the sessions were canceled. All children from our sample had experience in interacting with handheld devices, such as smartphones and tablets, while none of them were accustomed to IEs. Eight of them stated they had some limited experience with programming, i.e. they had already played a programming game or attended a robotics seminar at school, while regarding their experience with AR, six participants said that they had encountered AR books (with limited interactivity) a couple of times, while the remaining nine said that they had never heard of such technologies before.

Two specific individuals were appointed the roles of moderator and observer throughout the entire study to ensure that differences in personality traits or knowledge do not affect the study, while the moderator always used a written transcript to address the participants. Each study was scheduled to last about one hour (including time for introduction and debriefing), since children become fatigued after an hour of concentrated computer use [99].

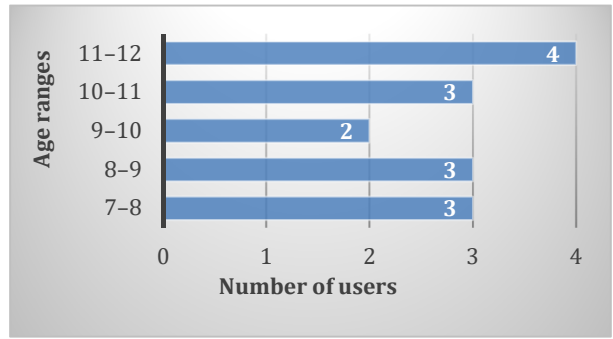

Figure 2: Age Distribution of the Participants

### 4.1 Procedure

**Pre-study**: As soon as the children, along with their parents, arrived at the facility, they were welcomed and introduced to the study moderator and the observer. Before entering the simulation space of the "Intelligent Living Room", the moderator offered a 15-minute tour of the facility (including other indicative simulation spaces that could spark their interest, such as an Intelligent Classroom and an Interactive Museum), explaining to the children the work that takes place in that environment. Next, the children were led to the living room of the Intelligent Home, and the moderator continued with a child-friendly demonstration of various aspects of that particular IE (i.e. the smart artifacts and their functionalities), in order to help them become familiar with the space. During both the tour and the demonstration of the living room, children were able to ask questions and interact with various systems so as to relax, start off easy and understand that unlike other similar places (e.g. laboratories, museums), interaction with the artifacts is not prohibited, but rather strongly encouraged. As a next step, the moderator explained the purpose of the study to the parents, asked them to sign the consent form, and requested that they remain silent during the study [99] and not intervene [104], so as to not introduce bias in the process. Then, the moderator engaged in a short casual conversation with the children, so as to help them feel comfortable, and reduce any anxiety, shyness, or uncertainty. During that conversation, it was communicated to the children that they had a very important role: "*to help the team understand whether they have indeed created a good game for children to program their surroundings*", and they were reassured that the more problems they uncover, the more helpful they become. Playing the role of the "not know-it-all" adult and putting the child in the shoes of the "expert" truly seemed to empower and motivate even the shyest of children. Finally, the moderator briefly explained the concept of MagiPlay, gave some examples of trigger-action rules, showed where the smart artifacts reside in the room, and concluded by informing the children that they could ask for a break or to stop at any time.

**During the study**: The children were asked to play four levels of the game (one introductory level and three levels of progressive difficulty), during which they were asked to create and deploy four (4) rules respectively, defining the behavior of artifacts inside the "Intelligent Living Room":

L1. (Training Level) When the phone rings, the speaker should turn off
L2. When the door closes, the lights should turn off
L3. When the door closes, the lights should turn on and the TV should turn on
L4. When mum returns from work and we sit on the sofa, the TV should play the movie "Cars II" and the lights should turn off.

At the end of each level, the moderator triggered the corresponding rule, so that the children could see the result of their effort materializing before their eyes; for example, at the end of Level 1, the moderator made a call to the IP phone in order to trigger the relevant rule.

The first level acted as a warmup introductory task that would get the participants comfortable with the User Interface (UI), build up their confidence, and increase their motivation through an "early success". Some guidance was provided in the form of narration and visual cues, in order to help children complete the first rule. For example, the command bricks that they had to collect were highlighted, while the agent was giving prompts such as "First you need to collect the artifacts that you want to program", and "What do you want to happen when the phone rings?". This interactive in-game tutorial aimed to address children's unfamiliarity with the game's fundamental concepts, namely programmable smart objects, creation of trigger-action rules, and interaction in AR. The moderator was responsible for orchestrating the entire process and assisting the children when required, i.e. offering them help in case they requested it or showed visible signs of confusion (e.g. remaining idle). At the same time, both evaluators were keeping notes, using a custom observation grid appropriate for recording a variety of metrics (e.g. interaction errors, hints given by the moderator), as well as children's comments, non-verbal behaviors, and emotions.

**Post-study:** After completing the four tasks (i.e. L1-L4), the children were asked to answer a few questions, in order to express their opinion on the system and the abilities of the IE, what they liked or disliked the most, and whether they had any suggestions to make. In the end, the moderator thanked the children, acknowledged how helpful they were, and offered them a small gift (e.g. pocket-sized board games) to express the gratitude of the entire team. Finally, the children had the opportunity to spend some extra time in the facility, and play with other interactive games that have been developed in-house in various domains (e.g. education, entertainment, arts).

### 4.2 Data Collection

During the study, we collected various data, which were subsequently examined by a team of experts including one HCI expert, two design experts for children, one interaction designer, and one psychologist.

**Pre-study questions.** Before the beginning of the experiment, the moderator, in an attempt to build up a trusting relationship between her and the children, engaged in a casual conversation, asking short questions such as: "what is your favorite super hero?" or "what is your best's friend name?". Amid these questions that the moderator asked so as to find a common ground with the child, there were four questions towards establishing the user's profile and facilitating the analysis of the results: (i) "How old are you?", (ii) "Do you play with your parents' tablet or smartphone?", (iii) "Have you ever played a programming game either computer-based or a board game?", and (iv) "Have you ever interacted with an application using AR technologies?".

**Qualitative Data.** The moderator, who was sitting next to the child, was keeping notes with the use of a custom observation grid appropriate for recording non-verbal child behavior (e.g. smiling, frowning) and emotions (e.g. happy, overwhelmed), as well as the child's comments. This type of data can reveal insightful information such as whether a child is enjoying the game, which parts of the process require the most effort to complete, and their overall feelings and attitude towards the game at key moments (e.g. when successfully completing a task or when facing difficulties).

**Quantitative Data.** The observer was sitting behind the child and was using equipment that permitted viewing the screen of the tablet (i.e. screen mirroring). Thus, the observer was able to record various information, such as interaction issues, errors during the rule creation process, and hints given by the moderator. The data concerning the interaction errors can help reveal whether a child experiences difficulty with the selected interaction modality (i.e. AR), while the information regarding the rule creation errors can reveal whether the block-based and trigger-action paradigm are suitable (i.e. supports children in identifying which facilities of an IE can be used and permits children to easily and correctly formulate the desired rule). Finally, observing whether a child required help for a specific task is vital for uncovering potential issues.

**Post-study Questions.** As soon as the session ended, the moderator asked the children six questions in order to understand their opinion regarding the "Intelligent Living Room", their attitude towards the game and its objective, whether they faced difficulties and if they had any suggestions for improving it. Since young children can experience difficulties in articulating their thoughts [104], and in order to avoid making them feel overwhelmed, they were asked for their opinion indirectly. For example, instead of asking what they did or what they thought of the game, we asked them to tell us how they would describe what they have done in the session to their best friend the next time they see them, or what their younger sibling(s) might think of the game.

### 4.3 Results

*4.3.1 Q1: Can children (aged 7-12) comprehend the concept of IEs?*

The concepts of IEs and Programmable Smart Objects were complex and unknown to all participants. However, our study revealed that through MagiPlay the children understood that they could program their surroundings in order to behave in a specific manner. The latter increases in significance when considering that, during the introduction, the moderator provided only a brief explanation of the concept and the capabilities of an IE and its artifacts. Children though, while playing, developed a good comprehension of what an IE is, what kind of functions it can execute to assist daily activities, and why it is characterized as "intelligent".

In more detail, examining the children's answers to the question: "*What would you tell your best friend that you accomplished via MagiPlay?*", we observed that the responses of 10 out of 15 children indicated that they apprehended that the rules were the means to dictate the behavior of the house (e.g. "*I was able to program the house to do the given four rules*", "*I created rules and the game executed them*"). Additionally, 4 children provided answers signifying that they understood that the user can be in control of the environment (e.g. "*The game enabled me to state what I want to happen when something else happens*", "*The game allows me to open or close something when I want to*"). Interestingly, one child (an 8,5-year-old girl) provided a more sophisticated answer, stating that "*I made artifacts to do actions without the need for me to tell them what to do*", which suggests that she comprehended the essence of end-user programming. Such responses can reveal that children have achieved the "Knowledge" and "Comprehension" level of intellectual behavior, described in Bloom's taxonomy[1] of educational objectives [105], while the fact that all children managed to complete Levels 1-4 with minimum help shows that they also achieved the "Application" and "Analysis" levels. Additionally, the children's answers to the question: "*What would you like to program in your Home if MagiPlay was installed in the family's tablet?*" revealed that the majority of them (11 out of 15) have also reached the levels of "Synthesis" and "Evaluation". In particular, most of the provided answers can be considered as valid examples of what an Intelligent Home would be capable of: "*I would like to program my home to prepare breakfast or lunch when I don't know what to eat*", "*When I wake up to turn on the TV*", "*When I want to study on my desk, I would program it to turn on my desk lamp automatically*", "*When I wake up, my breakfast is automatically prepared*". These responses indicate that children did not simply recall the rules they created while playing the game; rather, they were able to transfer the acquired knowledge to their own personal context.

### 4.3.2 Q2: Can children (aged 7-12) dictate the behavior of the IE by creating appropriate rules?

Children from the age of five are familiar with verbal conditional statements [34], hence, as expected, creating trigger-action rules proved to be intuitive and natural for all participants. When asked by the game to create a specific rule, children were able to verbally express, in various informal manners, their intentions on how they were about to proceed so as to build it. Most of the time, the expressions they used closely resembled the structure of trigger-action rules; this observation indicates that this programming paradigm is appropriate for children, since it matches their way of thinking. In particular, the findings of this study revealed that all children, after easily completing the first level, were able to continue with the subsequent levels without facing any major difficulties in terms of comprehending and formulating the expected conditional logic, while being offered minimum help from the moderator. It is worth mentioning that apart from the UI, all the command and literal bricks (e.g. WHEN, THEN, AND, "Turn On", "Turn Off") were also translated to the children's native language, which facilitated the selection of the correct ones, according to the hints provided verbally by the agent.

Our findings revealed that after the introductory level that accustomed the children to the adopted block-based programming paradigm, they were able to formulate the required rules easily. In more detail, during the second task, 80% of the children successfully completed the objective without making any error or receiving any assistance from the game or the moderator; the remaining 20% made 3 errors in total, mainly related to the placement of the required bricks (Table 1).

Table 1: Errors observed during the rule creation process

|  | **Error Description** | Lvl |
|---|---|---|
| 1 child | The child tried to place the artifact brick first and then the command brick, instead of the opposite. | L1 |
| 1 child | The child placed the artifact brick directly on the baseplate, instead of placing it on top of the command brick. | L2 |
| 1 child | The child did not use the command brick THEN. | L2 |
| 1 child | The child selected the wrong trigger on the literal brick. | L2 |
| 2 children | The child placed a wrong artifact brick on top of the command brick. | L3 |
| 13 children | The child did not understand that a THEN brick should follow the AND brick. | L3 |
| 2 children | The child forgot that a THEN brick should follow the AND brick. | L4 |

---

[1] Levels of intellectual behavior that are important for learning. 'Knowledge': Recall data or information. 'Comprehension': Understand the meaning of a problem, be able to translate into own words, 'Application': Use a concept in a new situation, 'Analysis': Split concepts into parts and understands the structure, 'Synthesis': Produce something from different elements, 'Evaluation': Make judgments, justify a solution, etc.

During Level 3, we observed a notable increase in the number of errors made (i.e. fifteen errors). This upturn is explained by the fact that Level 3 involved a concept that was not introduced to them beforehand: the creation of a rule with two actions (i.e. the lights should turn on AND the TV should turn on). Despite the fact that all children understood that they had to create a rule with multiple actions, 85% of them did not realize that an additional THEN brick should follow the AND brick. Children even disregarded the visual cues implying that nothing can be stacked on top of an AND brick (i.e. the top of the brick was flat with no visible connectors, while the only available connection points existed on its sides), and kept trying to place the artifact brick that corresponded to the second action on top of it. A possible explanation to that issue is that the expected way to formulate an expression describing that particular rule in natural language would be: "WHEN the DOOR closes, THEN the LIGHTS should turn on **AND** the TV should turn on", instead of: "WHEN the DOOR closes THEN the LIGHTS should turn on **AND THEN** the TV should turn on". Additionally, during Level 4, two children repeated the same mistake while building the trigger of the rule ("WHEN mum returns from work AND WHEN we sit on the sofa"), but as soon as the moderator provided a hint, they stated that they forgot to place the second WHEN brick. Possible solutions to tackle that issue would be to either automatically append a WHEN or a THEN brick next to an already placed AND, by checking if a trigger or an action should follow, or to completely eliminate the need for additional bricks. The errors recorded in Levels 1 to 4 are summarized in Table 1. A significant decrease in the errors in Level 4 can be observed, while no issues regarding children's ability to create trigger-action rules during the game were detected; the identified issues regarded the metaphors and specifics that MagiPlay employed for transferring the trigger-action paradigm in a block-based programming environment.

## 5 DISCUSSION

The results of this study not only indicated that children can indeed understand the concept and usefulness of IEs, but also that they are capable of programming their behavior, which they find interesting and enjoyable. In more detail, the 15-minute tour of the facility, the brief introduction provided by the moderator, and the interaction with the game, seemed to help the children develop a good comprehension of what an IE is, what kind of functions it can execute to assist daily activities, and why it is characterized as "intelligent". This reveals that in a relatively short time frame (less than an hour), children were adequately familiarized with many advanced concepts (e.g. programmable artifacts, trigger-action programming), and the overall paradigm of IEs that was completely unknown to them. Not only were they able to apprehend the idea of dictating the behavior of artifacts via rules, but they also learned how to formulate them quite quickly. The above validates the hypothesis that: since children (aged 7-12) are cognitively able to use logical thought [26] and learn the basic concepts of coding [37], they are also able to program the intelligent artifacts around them.

This is also confirmed by the comments they made after the end of the study, which highlight their ability to fabricate rich scenarios such as *"I would add more smart artifacts in order to be able to automate a lot of things, such as a moving staircase to reach books on higher shelves"*, or *"I would add more rooms, which I could program. I would like to set rules for an entire day"*, and *"I would make the system smart (e.g. when no one is in the room, then the lights should turn off)"*. Moreover, the participants of this study surpassed our expectations since older children (i.e. 12 years old) were capable of understanding inherently complex concepts regarding IEs and their manipulation. In particular, a child raised the issue of dealing with overlapping rules (i.e. conflicts) and wanted to know how the game could assist the user in managing them. Additionally, another 12-year-old child comprehended the limitless potentials of IEs and expressed his desire to create smarter rules, such as *"when no one is in the room, then the lights should turn off"*, which does not rely on simple instant events (e.g. the door opens), but rather exploits contextual information (i.e. user presence).

Thus, children within the age range 7-12 can and should be considered as potential users of end-user programming tools for IEs. As the moderator offered a short tour of the facility, we were able to observe children's reactions to this "entirely new world" full of possibilities unveiling before them. In more detail, their eyes were full of excitement, while they uttered exclamations of amazement (e.g. *"Wow"*, *"Cool"*) when interacting with various intelligent systems. The children's answers to the moderator's questions at the end of the study revealed that they enjoyed MagiPlay and had fun playing it. Some of the most noteworthy answers are: "*I really liked it, I have never seen a game like that*", "*It is very interesting and fun*", and "*I liked that it is an actual game that you make rules*". Moreover, children's reactions during the study revealed that their favorite part of the game was when they could actually see the rule being executed, thus affecting their surroundings. During this process, children exhibited signs of surprise followed by enthusiasm, which evinces that they enjoyed observing the realization of their work. Subsequently, such environments could provide the means through which children would be able to express their creativity by building innovative scenarios that support their daily activities, and configure their surroundings to match their taste in a ubiquitous manner.

## 5.1 Preliminary Implications Regarding the Design of Future Systems

During this study, despite the limited sample size, we were able to extract a number of preliminary implications regarding creating end-user programming tools for IEs targeting children aged 7-12 years, which are worth future research.

- **Trigger-action programming is an appropriate form of programming IEs**, since trigger-action rules seem to come naturally to children for dictating the behavior of smart artifacts. Literature has already demonstrated that even inexperienced users can successfully program using this paradigm [96]. Our study showed that this applies to children as well, and confirmed that since young children are familiar with verbal conditional statements [34], this form of programming would be easy to follow. More specifically, after completing the first level easily, all children were able to continue with the subsequent levels without facing any major difficulties in terms of formulating the expected conditional logic, while being offered minimum help from the moderator.

- **AR is a promising technology appropriate for direct interaction with the programmable artifacts and first-hand experience of the outcome of the created rules**. In more detail, since AR interfaces permit users to view and interact with virtual 3D objects superimposed in the real world, AR is an appropriate technology for (i) simulating the created rules in order to experience its effects, and for (ii) discovering and collecting smart artifacts within the physical environment. The latter assists children in better understanding which artifact's behavior they are going to manipulate, by allowing them to make direct connections between the real, physical objects and the 3D virtual blocks representing them. These findings are supported by research stating that AR increases task-related intuitiveness [90] by providing a more natural interface [106], through which users can interact straightforwardly with spatial information [107], while ensuring location awareness [91]. Thus, younger children benefit from having active experiences with manipulatives that promote the development of associations between concrete and symbolic representations [92]. On top of that, during the study, the majority of the children revealed that their favorite part of the game was the process of moving around the environment and collecting the smart artifacts.

- **Allowing children to explore their surroundings and actively move through the environment results in enjoyment and spatial awareness.** The notes kept by the observer revealed that the majority of the children were smiling and having fun during the collection of the different artifacts, which suggests that it was a pleasant and entertaining process. The latter is further corroborated by the children's responses; both their comments during the study itself and their answers during the final debriefing. Such two representative statements are a child mentioning "*I really liked that I had to move in the room and collect the artifacts*", while another one saying "*I liked the idea of collecting the actual artifact*". In any case, several studies from the '80s until today [28–31], have concluded that children construct much of their knowledge through active manipulation of the environment.
- **Gamification elements are beneficial for enhancing children's engagement and creating a positive and stimulating atmosphere in this setting**. This is corroborated by the fact that the moderator and observer did not notice any direct or indirect signs of boredom, frustration, or indifference, while 70% of the children requested to play the game again after the end of their session. Additionally, 9 out of 15 children stated that they liked the fact that they could 'unlock' artifacts, so as to use them in the rule creation process, and 6 out of 15 expressed their interest in finding out the 'experience points' that they had collected.

## 5.2 Suggestions for Studies with Children in Intelligent Environments

Regarding studies with children in simulation spaces or existing IEs, our findings suggest that **a child-friendly demonstration of various aspects of the IE helps the child become familiar with the intelligent space and its capabilities.** Presenting the intelligent space where the study will be hosted, or of any other collocated intelligent facilities, even if they are irrelevant to the system under study, seems to be beneficial. Permitting children to explore the space, ask questions and interact with various technological components assists them in getting acquainted with the concept and therefore feel safe and comfortable. Additionally, if the moderator is the one giving the demonstration, they have the opportunity to bond with the children and build trust between them. Similarly, **allowing children to spend some additional time inside the facility after the study ends, might lead to uncovering their opinion regarding the system.** Making the child feel relaxed after the study might motivate them to expose information to their parents/siblings or the moderator regarding the system. For example, a child asked his mother *"When is it going to be available on the Appstore?"* and another said, *"The game wasn't difficult"*. Despite the fact that there was a query in the "Post-study Questions" asking *"Can you use your imagination to tell me what you would do to make the game better?"*, some of the children responded *"I don't know"*; however, while interacting with other intelligent systems of the facility, they retrospectively addressed the moderator and made suggestions for improving MagiPlay.

## 6 LIMITATIONS AND FUTURE WORK

A limitation in our study is that it was conducted in a controlled environment (simulation space), and not "in the wild", in the actual homes of the children-participants; our future plans include a study in real-life contexts. Moreover, while we had originally recruited more than forty-five participants, we only had the opportunity to complete fifteen sessions, due to the COVID-19 pandemic. Nevertheless, we managed to gather various insights regarding the ability of children (aged 7-12) to understand the nature of IEs and their dexterity in creating rules that dictate their behavior. Interestingly though, after the 10th user, the number of newly identified issues was rather small, as we observed the same findings repeatedly. Nonetheless, our future goals include observing more users during the creation of rules, including additional logical operators (i.e. OR, NOT) in both parts of the rule (i.e. trigger and action). We also wish to explore rules that entail artifacts residing in different rooms of a house, and investigate how this could affect the interaction paradigm of the game, e.g. whether children will be willing to move between different rooms to collect the necessary artifacts. Additional future plans include studies focusing on children aged 10-12, since our findings indicate that they are capable of understanding more complex concepts regarding the programming of IEs. To this end, we wish to investigate their response to even more sophisticated notions, such as priority of rules, potential conflicts, and multi-user scenarios.

Moreover, since MagiPlay was the means to explore the children's attitude towards programming IEs, it is not apparent which of the selected game mechanics (AR, gamification, trigger-action programming, and building blocks) mostly contributed to the positive results; hence future research could examine which of the aspects played a prominent role, and to what extent. For example, it would be interesting to investigate whether children would still find it fun and engaging to program the IE if this was not performed via a game (without the aspect of gamification), or if they would have trouble comprehending the aspect of IEs without AR, which now arguably bridges the virtual world of programming with the real world of triggers and actions and artifacts.

An interesting question that should be further researched is the extent to which children should be allowed to manipulate their environment. In particular, there are various pitfalls in letting children come up with their own rules; for instance, a child could create rules that turn on the lights or sound alarms in the middle of the night. Thus, this is a current limitation that needs to be addressed, e.g. creation of a parental supervising mode. In any case, smart objects of the future should be "child-proof", in the sense that they know when they are being manipulated by children, so as to expose a different interface and restrict their capabilities to ensure safety.

Finally, in a broader context, research could explore whether systems like MagiPlay can be employed by educators as learning tools to teach children how to program their intelligent surroundings, e.g. in a robotics class in today's school context, or AR/VR courses in the classrooms of the near future.

## 7 CONCLUSION

This paper presented a study (N=15) with children aged 7-12 aiming to determine whether (i) children understand the concept of Intelligent Environments and (ii) can program the smart artifacts they include. To that end, the children-participants played four levels of an AR serious game (called *MagiPlay*), allowing to create trigger-action rules to dictate the behavior of IEs. In particular, we addressed research questions concerning (i) children's comprehension of the concept of IEs, and (ii) children's ability to dictate the behavior of an IE by creating block-based, trigger-action rules. Moreover, interesting insights were gathered with respect to children's interaction with and within IEs and user studies with children participants taking place in IEs.

The present study showed promising and positive results regarding children's interaction with the game, and thus with systems exhibiting such qualities and employing concepts such as AR, 3D programmable blocks, trigger-action-like rules, and IEs. Our findings showcased the children's excitement in using MagiPlay, as well as their ability to effectively employ it and achieve the tasks that were asked of them, successfully managing to program their intelligent surroundings. We hope that our findings can serve as a basis for future research on children's interaction with Intelligent Environments, and in particular with regards to children programming the smart artifacts they encompass, as they will be the future users and adopters of such spaces.

## ACKNOWLEDGMENTS

This research is supported by the FORTH-ICS internal RTD Program 'Ambient Intelligence and Smart Environments'.

## SELECTION AND PARTICIPATION OF CHILDREN

The study and its protocol were approved by the Ethics Committee of our research center. The study was explained to the children's legal guardians/parents, who gave their informed consent for inclusion before participation. At the start of each session, the researcher briefly explained the study to the child and asked for a verbal agreement to participate, informing them that if they wanted to stop for any reason, they would stop immediately and exclude the data from the study. A total of fifteen children took part in the study (7 girls, 8 boys). The children were invited through personal contacts of the research team (children of colleagues, classmates of their children). All participants' personal data were stored securely, and all personally identifiable data were removed.